\documentclass[default,iicol]{sn-jnl}

\usepackage[utf8]{inputenc}

\jyear{2021}%

\theoremstyle{thmstyleone}%
\theoremstyle{thmstyletwo}%

\theoremstyle{thmstylethree}%

\usepackage{graphics}

\DeclareUnicodeCharacter{2212}{-}
\begin{document}

\title[Article Title]{Ultrafast racetrack based on compensated Co/Gd-based synthetic ferrimagnet with all-optical switching}


\author*[1]{\fnm{Pingzhi} \sur{Li}}\email{p.li1@tue.nl}

\author[1]{\fnm{Thomas J.} \sur{Kools}}

\author[1]{\fnm{Bert} \sur{Koopmans}}

\author[1]{\fnm{Reinoud} \sur{Lavrijsen}}

\affil*[1]{\orgdiv{Department of Applied Physics}, \orgname{Eindhoven University of Technology}, \orgaddress{ \city{Eindhoven}, \postcode{5612 AZ},  \country{Netherlands}}}








\maketitle
\newpage
\textbf{Spin-orbitronics \cite{Kim:2022ua, Dieny:2020ub} and single pulse all-optical switching (AOS) \cite{Kimel_AOS_Review2019} of magnetization are two major successes of the rapidly advancing field of nanomagnetism in recent years, with high potential for enabling novel, fast and energy-efficient memory and logic platforms. 
Fast current-induced domain wall motion (CIDWM) \cite{Caretta:2018aa} and single shot AOS \cite{Ostler:2012aa} have been individually demonstrated in different ferrimagnetic alloys.
However, the stringent requirement for their composition control \cite{Beens:2019aa,Ostler:2012aa} makes these alloys challenging materials for wafer scale production \cite{Kimel_AOS_Review2019}. 
Here, we simultaneously demonstrate fast CIDWM and energy efficient AOS in a synthetic ferrimagnetic system based on multilayered [Co/Gd]$_2$. We firstly show that AOS is present in its full composition range.
We find that current-driven domain wall velocities over 2000 m/s at room temperature, achieved by compensating the total angular momentum through layer thickness tuning. Furthermore, analytical modeling of the CIDWM reveals that Joule heating needs to be treated transiently to properly describe the CIDWM for our sub-ns current pulses. Our studies establish [Co/Gd]-based synthetic ferrimagnets to be a unique materials platform for domain wall devices with access to ultrafast single pulse AOS.}

Research in spintronics over the last decades has demonstrated a possibility for further evolution of solid-state electronic application platforms beyond CMOS.
The racetrack memory \cite{Parkin:2008aa}, an envisioned novel data storage device concept based on using magnetic domain walls (DWs) \cite{KUMAR20221}, utilizes frontier spintronic mechanisms to function as an ultra-dense on-chip memory with an operation speed comparable to low-level cache \cite{Blasing:2020aa}. 
The efficiency of racetrack memory relies on the fast CIDWM in a material with perpendicular magnetic anisotropy. 
The maximum velocity was significantly enhanced over the years by combining spin-orbit torques (SOTs) and the Dzyaloshinskii-Moriya interaction (DMI) \cite{Miron:2011ab,Ryu:2013aa,Haazen:2013aa} with synthetic antiferromagnets (SAFs), which resulted in reported DW velocities close to 750 m/s \cite{Yang:2015aa}. 

Despite the large improvement, the energy efficiency is still limited due to the weak strength of the antiferromagnetic (AF) coupling. Therefore, the materials platform of rare earth (RE)-transition metal (TM) compounds garnered considerable attention, promising faster CIDWM due to the much stronger direct AF coupling than the indirect exchange coupling \cite{Parkin:1991aa} utilized in SAFs. Furthermore, the SOTs used to drive the DW promise to be highly efficient in the RE-TM systems as a result of the long spin coherence length \cite{Yu:2019vn}. Cosequently, high velocity CIDWM has been reported in Co-Gd-based ferrimagnetic alloy systems \cite{Caretta:2018aa,Okuno:2019aa} when the angular momentum in the magnetic material is compensated, being at least a factor of three faster than that of the previously reported SAFs. 

Besides the efficient CIDWM, single pulse all-optical switching (AOS) of the magnetization \cite{Ostler:2012aa} in the RE-TM systems has obtained significant attention thanks to its sub-picosecond \cite{Radu:2011aa} energy efficient \cite{Khorsand:2012aa,Kimel_AOS_Review2019,Li:2021wr} magnetization switching enabled by the ultrafast angular momentum transfer upon laser excitation \cite{Mentink:2012aa}. 
This can be useful as a new generation of ultrafast magnetic memory, as well as a data buffer between electronics and integrated photonics \cite{Becker:2020aa,Sobolewska:2020aa,Kimel_AOS_Review2019}. Recently, a synthetic ferrimagnetic system based on a Pt/Co/Gd \cite{Lalieu:2017aa,Li:2021wr} layered structure has shown high robustness \cite{wang2020picosecond} for such a hybrid integration.

These kinds of synthetic ferrimagnets have some distinct advantages over RE-TM alloys. For instance, AOS is not limited by the exact composition \cite{Beens:2019ab}. They also withstand thermal annealing \cite{Wang:2020ab} and offer easier magnetic composition control at wafer scale than the alloy system, as well as better access to interface engineering. Therefore, it has been proposed that such a materials platform has high potential to realize a hybrid integration of DW memory in photonic platforms to further enhance their storage density \cite{Lalieu:2019aa, Becker:2020aa}.


So far, the CIDWM of Co/Gd bilayers \cite{BlaesingECT2018,Lalieu:2019aa} has been investigated. However, the highest reported velocity, achieved at cryogenic conditions \cite{BlaesingECT2018}, was several times lower than that reported in alloys \cite{Caretta:2018aa}, in part due to large net angular momentum, low compensation temperature as well as DW pinning effects. In this report, we therefore propose a materials platform based on the [Co/Gd]$_2$ synthetic ferrimagnet capable of accommodating both efficient CIDWM at room temperature (RT) and single-pulse AOS, with compatibility for wafer scale production as well as robustness for engineering. 
 

\begin{figure*}[h!]
    \centering
    \includegraphics[width=1\linewidth]{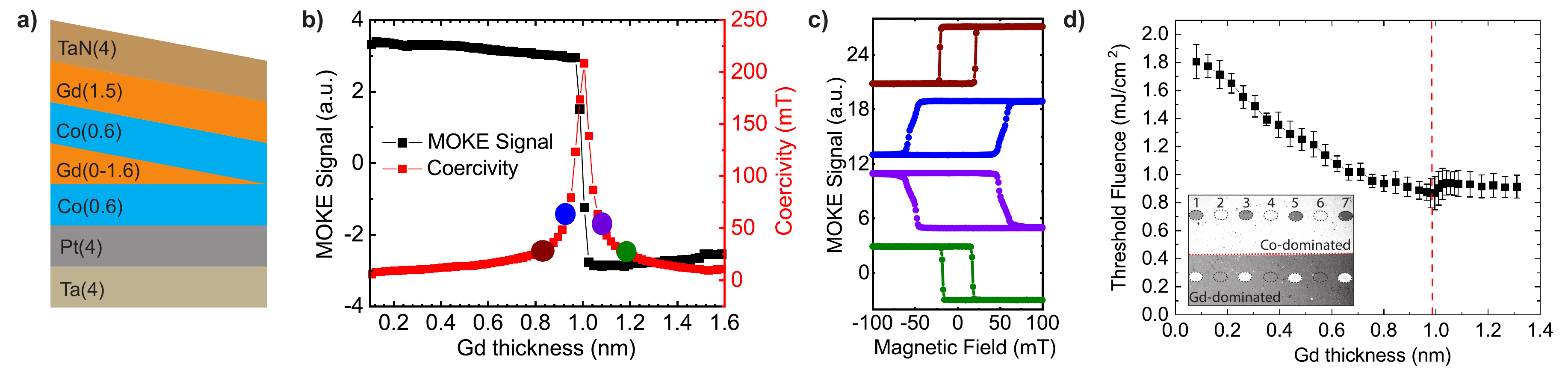}
    \caption{a): Schematic illustration of the layer stack used in our study (thickness in nm). b): Result of static MOKE study, where the MOKE signal (black) is defined by the difference in signal intensity between saturation at positive and negative applied field. Coercivity (red) was measured at a fixed scanning speed of 10 mT/s (red). c): Hysteresis loops measured at the sample thicknesses marked by the coloured dots in b). d): Threshold fluence for all-optical switching across the wedge shown in a). Inset shows toggle switched domains at two sides of the compensation thickness for varying amount of subsequent pulses.}
    \label{fig:FigureStatics}
\end{figure*}


The magnetostatic and AOS properties of the [Co/Gd]$_2$ materials platform under investigation in this work are shown in Fig. \ref{fig:FigureStatics}. A Ta/Pt/[Co/Gd]$_\text{2}$/TaN multilayer is deposited on a Si/SiO$_\text{2}$ substrate, where the first Gd layer is wedged (see Fig \ref{fig:FigureStatics}.a and Methods) to tune the net magnetic moment. Due to the proximity effect by the Co, a net magnetization is induced in the Gd which drops off steeply away from the Co interface \cite{Nesbitt:1962vk}.
Compared to the bilayer Co/Gd \cite{Lalieu:2017aa,BlaesingECT2018}, we double the magnetic volume of the Co in [Co/Gd]$_\text{2}$, while tripling the number of interfaces at which magnetization is induced in the Gd. This allows us to more easily compensate both the angular momentum and magnetic moment of the system at RT by varying the individual layer thicknesses.
  
We measured hysteresis loops by polar MOKE at RT across the thickness range of Gd between 0-1.6 nm. In Fig. \ref{fig:FigureStatics}.b we plot the MOKE signal intensity, defined as the difference in signal between positive and negative saturation, as well as the coercivity obtained from these hysteresis loops. The loops corresponding to the colored dots in Fig. \ref{fig:FigureStatics}.b are presented in Fig. \ref{fig:FigureStatics}.c. 

The 100\% remanence indicates a well-defined perpendicular magnetic anisotropy as well as tight exchange coupling between the layers. 
Importantly, at the Gd thickness of 0.97 nm, the MOKE signal switches sign (see black curve in Fig. \ref{fig:FigureStatics}.b). This can be observed as well in Fig. \ref{fig:FigureStatics}.c as an inverted hysteresis loop. 
At this thickness the magnetization is compensated, i.e. the Co and Gd magnetization cancel each other. 
This is further evidenced by the divergence in coercivity, and confirmed by superconducting quantum interference device measurements (see Sup. \ref{section:SQUIDcharac}). Thus, upon increasing the thickness of Gd, the magnetic balance is shifted from being Co-dominated to Gd-dominated. 

With respect to the magnetostatics, we note that the magnetization and angular momentum compensation thickness in these [Co/Gd]-based ferrimagnets are similar but not identical due to the different Landé g-factors of Co ($\sim$ 2.2) and Gd ($\sim$ 2.0). Therefore, when discussing compensation without further specification in the remainder of this text, we refer to angular momentum compensation as this is the relevant condition for efficient CIDWM \cite{BlaesingECT2018,Caretta:2018aa,Kim:2017aa}.

To investigate whether our 4-layer Co/Gd system is all-optically switchable, and how it depends on compensation conditions, we performed AOS experiments by illuminating the wedge with single femtosecond laser pulses with varying pulse energy, and examined the sample under a polar Kerr microscope. We find that single pulse AOS is present at every thickness on the [Co/Gd]$_2$ wedge. An example of the resulting toggle switched domains around the compensation boundary is shown in the inset of Fig. \ref{fig:FigureStatics}.d, which can be observed from the contrast inversion (see Sup. \ref{AOSsupp} for more images). This distinct feature that the AOS occurs both in the deep Co-dominated and Gd-dominated composition is unlike that of alloys \cite{Khorsand:2012aa} and Co/Tb multilayers \cite{Aviles-Felix:2019aa}, in which AOS is only present within $\pm$1.5 $\%$ of the composition window. This shows the flexibility of our system in terms of AOS engineering.  

We further plot the threshold fluence as a function of Gd thickness, calculated from the switched area and the corresponding pulse energy \cite{Lalieu:2017aa}, in Fig. \ref{fig:FigureStatics}.d. We find that the threshold fluence depends significantly on the material composition, and is at a minimum around the compensation point. Initially, the threshold fluence decreases monotonously with Gd thickness up to 1 nm, with a reduction of the threshold fluence of more than 50\% reaching 0.9 mJ/cm$^2$, which corresponds to 25 fJ for a 50$^2$ nm$^2$ domain. Such a switching energy is an order of magnitude lower than that of typical SOT and spin-transfer torque switching of a ferromagnet \cite{Yang:2017aa}.   

 Although a full dissection of the exact mechanism behind the thickness-dependence of the threshold fluence is beyond the scope of this work, we note that previous studies have shown that reducing the Curie temperature of layered systems can lead to a reduction of the threshold fluence \cite{Beens:2019aa,Beens:2019ab,Lalieu:2017aa,Li:2021wr}. In our case, as the Gd layer increases, the Curie temperature of the total stack is expected to drop, because the mutual exchange stabilization of the two Co layers is weakened, leading to a reduction of threshold fluence.


\begin{figure*}
    \centering
    \includegraphics[width=1.0\linewidth]{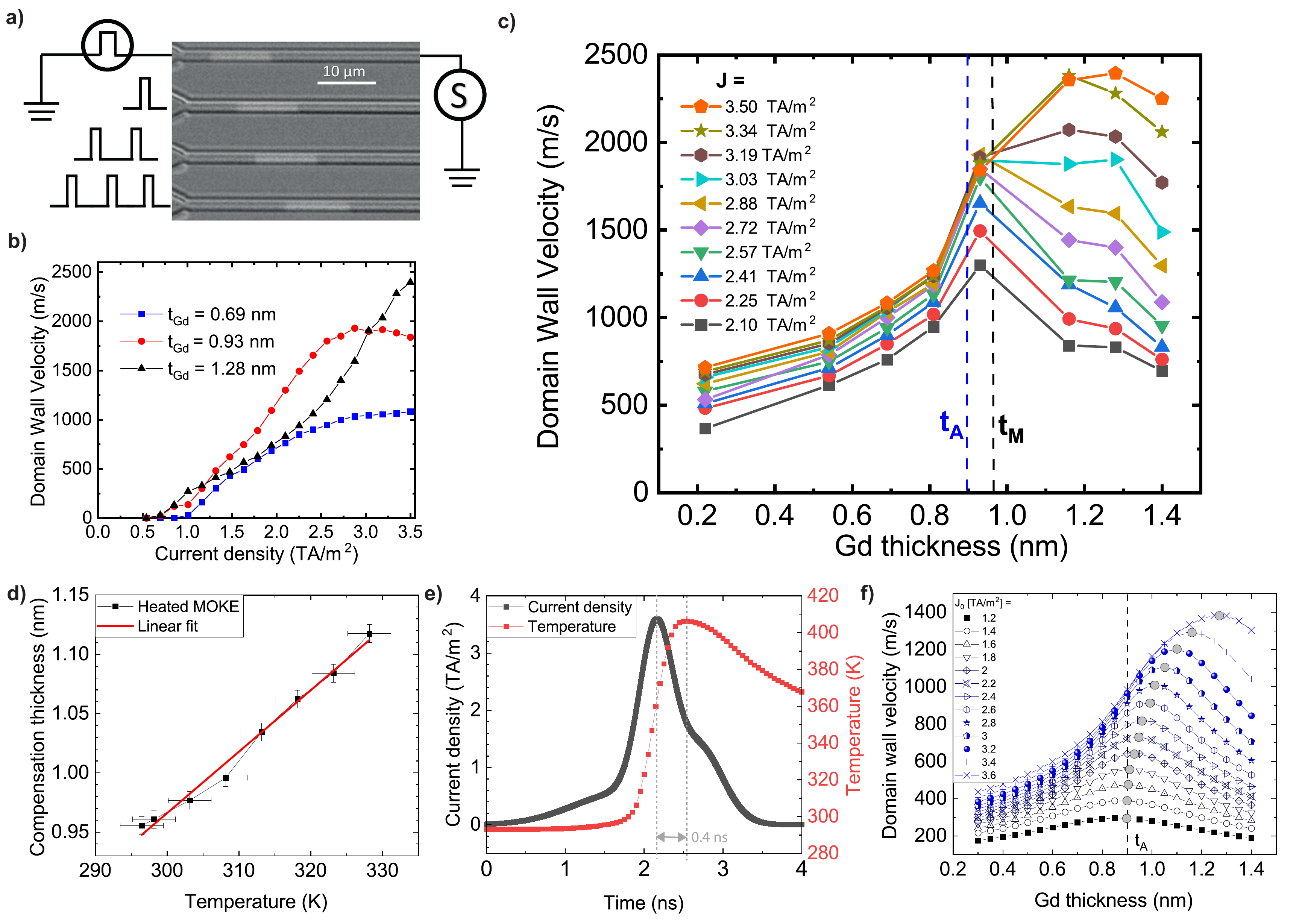}
    \caption{a): Example image from DW motion experiments performed by differential polar Kerr microscopy, where the setup is illustrated schematically as a pulse generator and a 6 GHz bandwidth oscilloscope. The DW position after applying consecutive pulses is shown. b): The measured DW velocity as a function of current density for three Gd layer thicknesses. c): The DW velocity as a function of Gd layer thickness for various current densities, where 10 \% error bars are not shown. Here we marked the quantified magnetization compensation thickness $t_\text{M}$ obtained from polar MOKE measurements as well as the corresponding angular momentum compensation thickness $t_\text{A}$ based on the estimation from magnetometry measurement. d): Compensation thickness as obtained from polar MOKE measurements of the [Co/Gd]$_2$ at various ambient temperatures. e): The temporal profile of the current pulse used in the experiment (normalized by the peak current density), and its subsequent temperature rise (as simulated with COMSOL) at 3.6 TA/m$^2$. f): Result of the theoretical 1-D modeling: the DW velocity as a function of Gd thickness, where the magnetization profile as well as its temperature dependence is in line with experimenes, while the current pulse and temperature profile follow the profile given in d). The velocity maximum is marked with a grey dot. The vertical dashed line indicates the angular momentum compensation thickness. 
    }
    \label{fig:DomainWallVelocity}
\end{figure*}
With compensation and AOS at RT confirmed, in the remainder of this work we demonstrate and explain the DW velocities of over 2000 m/s, as achieved in [Co/Gd]$_2$. In the case of such an AF coupled system with DMI, two contributions mainly drive the coherent motion of the DW: the DMI-torque and the exchange coupling torque. These two torques give rise to spin dynamics upon excitation by the spin Hall-effect-induced spin accumulation, and both lead to DW motion along the current direction (for positive DMI and spin Hall angle \cite{Miron:2011aa,Ryu:2013aa}).

It is well understood that the exchange coupling torque only drives the DW motion efficiently near the compensation point \cite{Yang:2015aa,BlaesingECT2018}. Adding additional AF coupled layers to a ferromagnet can facilitate this, however, can also increase the angular momentum of the DW to be translated.
Nonetheless, following the discussion in Sup. \ref{velocity enhancement}, we still expect a significant increase of the DW velocity in our study, as the enhancement of exchange coupling torque due to compensation significantly overwhelms the effect of added magnetic inertia. 

For this reason, we experimentally determined the DW velocity from the CIDWM in [Co/Gd]$_2$ as a function of lower Gd thickness. This study is performed in the same stack showcased in Fig. \ref{fig:FigureStatics}.a.
The wedged thin film was patterned into micro-structures as shown in Fig.  \ref{fig:DomainWallVelocity}.a (for details see Methods and Sup. Fig. \ref{fig:supJoule}.b). The magnetic domains were imaged by polar Kerr microscopy (see Fig. \ref{fig:DomainWallVelocity}.a).
Short current pulses of around 1 ns duration with varying amplitudes are injected into the structures to determine the DW velocity (see Sup. \ref{appendixsection1}). We observe that the DW moves coherently along the current direction (see Fig. \ref{fig:DomainWallVelocity}.a).
This confirms a left-handed Néel-type DW in our system as a result of a positive DMI, and a positive spin Hall angle \cite{Miron:2011aa, Ryu:2013aa}, as is characteristic of the Pt/Co interface \cite{Ryu:2013aa}.

In Fig. \ref{fig:DomainWallVelocity}.b we compare the resulting DW velocity as a function of peak current density for a Co-dominated (0.69 nm), compensated (0.93 nm) and Gd-dominated stack (1.28 nm), considering the magnetic composition at RT. 
We demonstrate that DW velocities of over 2000 m/s can be achieved at the highest current densities, at least three times larger than that observed in SAFs \cite{Yang:2015aa}. 

Furthermore, we notice from the crossing point between the red and black curve that the Gd thickness at which the highest velocity is observed becomes larger with increasing current density. We attribute this to the effect of Joule heating induced by the current pulses.
Earlier studies show that the Gd magnetization is quenched more severely during Joule heating than that of Co \cite{Nesbitt:1962vk, Svalov:2006aa,BlaesingECT2018,Cai:2020aa}. The consequence is that at higher temperature relatively more Gd than at RT is needed to achieve compensation. This fact is confirmed for our stacks by the upward shift of the compensation thickness with increasing sample temperature using polar MOKE (see Fig. \ref{fig:DomainWallVelocity}.d). We find that this hypothesis of a Joule-heating-induced shift of the compensation thickness explains the CIDWM experiments well. To understand this, it useful to consider two regimes in Fig. \ref{fig:DomainWallVelocity}.b in more detail.

For $J < 2.3$ TA/m$^2$, before saturation effects start to occur and for limited heating, we observe a steady rise of the velocity, consistent with earlier work on CIDWM in materials with AF coupling. As expected, the compensated sample is driven most efficiently. We visualize this further by plotting the DW velocity as a function of Gd thickness across the Co and Gd-dominated regimes, as is shown in Fig. \ref{fig:DomainWallVelocity}.c. Here, we observe that the DW velocity in this range of current densities spikes at the RT compensation thickness, further proving the origin of the velocity increase is due to the enhanced exchange coupling torque.

In contrast, for $J > 2.3$ TA/m$^2$, the DW velocity saturates both for the Co-dominated and the compensated sample. This is typical behaviour of the CIDWM of a ferromagnet \cite{Miron:2011aa,Ryu:2013aa, Yang:2015aa, BlaesingECT2018, Martinez:2014aa}, which is limited by the DMI-torque. Contrarily, the Gd-dominated sample shows a consistent velocity rise typically associated with a compensated system \cite{Yang:2015aa}. This is consistent with the hypothesis that the increased Joule heating shifts the compensation point to larger Gd thicknesses during the pulse, and is supported by the velocity peak shift observed in Fig. \ref{fig:DomainWallVelocity}.c.

In order to quantify the effect of the change in compensation thickness due to Joule heating on the CIDWM, we first numerically investigate the temperature transient for the current pulses applied in the experiment (see Sup. \ref{appendix:Joule heating}). In Fig. \ref{fig:DomainWallVelocity}.e we plot a typical current density profile and the corresponding modelled temperature change in the device. We observe a delay of $\sim 0.4$ ns between the peak current density, and the peak temperature. This is a consequence of the fact that the pulse duration approaches the characteristic time scale of the heat transport between the sample and the substrate (See Sup.\ref{appendix:Joule heating}). 


Importantly, around the peak current density where the driving torque exerted on the DW is largest, the sample temperature is changing rapidly from 0 to 50 $\%$ of maximum temperature within 0.4 ns. 
Therefore, the net angular momentum changes rapidly during the application of the pulse. This makes assigning an effective temperature during DW motion like previous studies \cite{Caretta:2018aa, BlaesingECT2018, Siddiqui:2018aa, Cai:2020aa} inaccurate.  

Therefore, we incorporated the transient current density profile, and the corresponding temperature rise into an analytical 1-dimensional model of the CIDWM (see Sup. \ref{Appendix: model description}), where we introduce a thickness and temperature dependent magnetization profile following our SQUID measurements (see Sup. \ref{section:SQUIDcharac}). The resulting modelled DW velocity as a function of thickness and current density is plotted in Fig. \ref{fig:DomainWallVelocity}.f. 

We find that the velocity peak shift with increased current density as shown in Fig. \ref{fig:DomainWallVelocity}.b is well described by our model. Furthermore, we also find that the rapidly changing angular momentum during the current pulse gives rise to a broadening of the velocity peak compared to the case with constant temperature previously described in \cite{Caretta:2018aa,BlaesingECT2018}(see Sup. \ref{Motivationdyntemp}).

These results suggest that in general transient treatment of the temperature profile is necessary to properly address the DW dynamics for ultrashort current pulses. Furthermore, to benefit from large exchange coupling torque at elevated temperature, which is a likely scenario in high-speed electronics, pushing the composition into the Gd-dominated regime is required.   


In summary, we have presented a materials platform of a synthetic ferrimagnet based on [Co/Gd]$_2$ which can be effectively tuned near the compensation point via layer thickness, and exhibits efficient single pulse AOS.
We also experimentally and numerically showed that fast CIDWM with velocities over 2000 m/s can be achieved with the help of angular momentum compensation.
Finally, we addressed the transient Joule heating effect during CIDWM, which gives important insight into our experimental observations and potential technological implementation. Our results therefore lay the foundation for an effective materials platform that exhibits both ultrafast CIDWM and AOS, paving the way for synthetic ferrimagnets to become the new paradigm of ferrimagnetic spintronics and providing a jumping board for further integration between photonics and spintronics.


\backmatter

\bmhead{Methods}

Magnetic films were deposited on Si substrates with a 100 nm thermally oxidized SiO$_2$ layer by D.C. magnetron
sputtering in a system with a base pressure of $\sim 5\times10^{-9}$ mBar. A thickness wedge is created with the help of a moving wedge shutter during sputtering. From these
thin films, nanostrips were fabricated using electron-beam lithography and
lift-off. The gold contacts were made by wire bonding. The out-of-plane component of the magnetization (M$_z$) of the
nanostrips was measured by polar Kerr microscopy. An in-plane field of 200 mT along the current direction, combined with a 3 TA/m$^2$ current pulse was used to nucleate domains. Both MOKE and Kerr Microscopy characterization were performed using a 700 nm continuous wave laser in the polar configuration. Here, we chose the wavelength of the laser light (around 700 nm) such that the detected signal is only sensitive to the magneto-optical response from the Co layer and Kerr rotation is maximized.

\bmhead{Supplementary information}

All supplementary information except for digital files are presented in the Appendix.

\bmhead{Contribution}
P.L. and T.J.K. contributed equally to this work in terms of design and conduct of the project. B.K. and R.L. supervised the project. All authors contributed to the writing of the manuscript.

\bmhead{Acknowledgments}

This  project  has  received  funding  from  the European Union’s Horizon 2020 research and innovation programme under the Marie Sklodowska-Curie grant agreement No.860060. This work is also part of the Gravitation programme ‘Research Centre for Integrated Nanophotonics’, which is financed by the Netherlands Organisation for Scientific Research (NWO).

\bmhead{Competing Interests}
Authors declare no conflicts of interest.

\begin{appendices}

\section{SQUID characterization}\label{section:SQUIDcharac}
\begin{figure*}
    \centering
    \includegraphics[width=\linewidth]{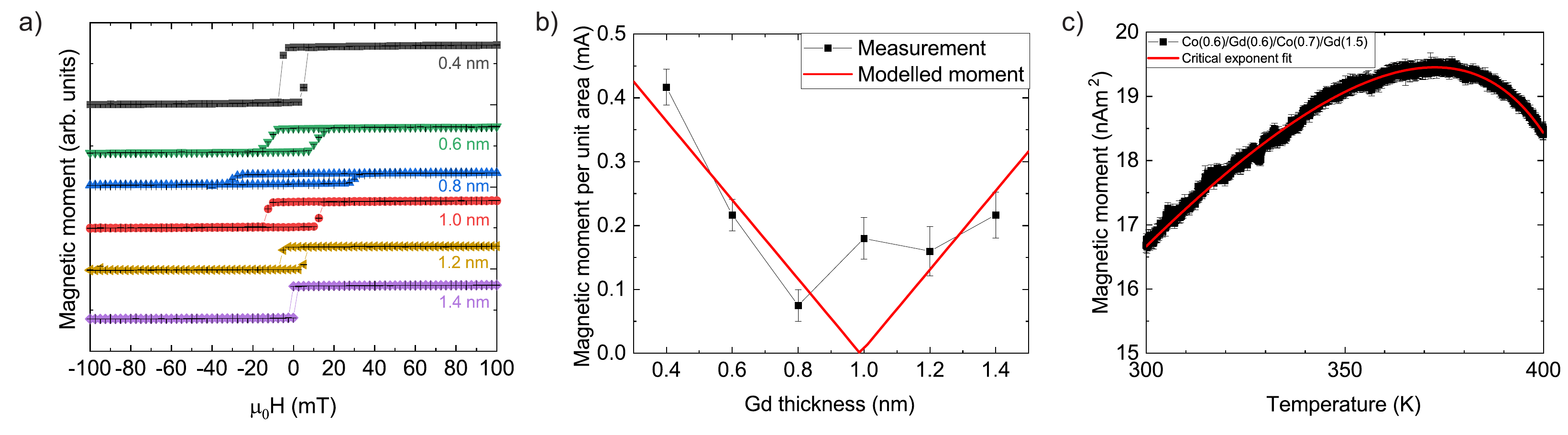}
    \caption{VSM-SQUID characterization of the out-of-plane moment of the Co(0.6)/Gd(x)/Co(0.7)/Gd(1.5) material system. a): Hysteresis loops across the compensation point at room temperature. b): Room temperature magnetic moment per unit area as a function of middle Gd thickness. c): Typical temperature dependence of the magnetic moment in a Co-dominated quadlayer sample. The red line indicates the best fit of equation \ref{equation:critexp}.}
    \label{fig:supSquid}
\end{figure*}
In order to characterize the magnetostatic properties of the sample, a series of 6 samples of composition TaN(4)/Pt(4)/Co(0.6)/Gd($t_\text{Gd}$)/Co(0.7)/Gd(1.5) /TaN(4), with $t_\text{Gd}=\left[0.2, 0.4, .. , 1.2, 1.4\right]$ nm were grown. Using VSM-SQUID, out-of-plane hysteresis loops were measured, which are shown in Fig. \ref{fig:supSquid}.a. The decreasing moment when going from a middle Gd thickness of 0.4 to 1.4 nm, and the diverging coercivity at the minimum of the measured moment are indicative of a compensated system. This can also be seen in the plot of the magnetic moment normalized by the sample area in Fig. \ref{fig:supSquid}.b. The temperature dependence of the magnetization was also investigated, and a typical result is shown in Fig. \ref{fig:supSquid}.c. In order to estimate the Curie temperature of the stack and obtain an expression for the magnetization $M_\text{tot}$, we follow earlier work on 3d-4f ferrimagnetic alloys in fitting a critical exponent to the data of the form \cite{CriticalExponentGd1,Kim:2017aa} given by:
\begin{multline}\label{equation:critexp}
     M_\text{tot}= M_\text{Co}(T)-M_\text{Gd}(T)=\\ M_\text{Co0}\left(1-\frac{T}{T_\text{C}}\right)^{\xi_\text{Co}}-M_\text{Gd0}\left(1-\frac{T}{T_\text{C}}\right)^{\xi_\text{Gd}},
\end{multline}
 
where $M_\text{Gd0}$ and $M_\text{Co0}$ and $\xi_\text{Gd}$, $\xi_\text{Co}$  are the magnetization at zero temperature and the critical exponent of Co and Gd, respectively. We assume $M_\text{Co0}=1.4$ MA/m, and find that the critical exponents $\xi_\text{Co}$= 0.50, $\xi_\text{Gd}$= 0.72, and $T_\text{C}$= 450 K describe the temperature dependence of the magnetic moment well (Fig. \ref{fig:supSquid}.c). We note that the magnetization in the Gd is an ill defined quantity, since the Gd transitions from partially magnetized due to the proximity effect, to truly ferromagnetic at low temperature. We therefore estimate the value of $M_\text{Gd0}$ by considering the known thickness dependence of the magnetic moment in the stack $m_\text{tot}$. This quantity is calculated as the sum of that of the individual layers as follows:

\begin{multline}\label{totmag}
    m_\text{tot}=m_\text{Co1}+m_\text{Gd1}+m_\text{Co2}+m_\text{Gd2} \\= M_\text{Co}(T)\left(t_\text{Co1}+t_\text{Co2}\right)-\\M_\text{Gd}(T)\left(2 t_\text{Gd1}+t_\text{Gd2}\right),
\end{multline}
where:
\begin{align}
m_\text{Co1} &=  M_\text{Co}(T)t_\text{Co1} \nonumber \\
m_\text{Co2} &=  M_\text{Co}(T)t_\text{Co2} \nonumber \\
m_\text{Gd1} &= 2 M_\text{Gd}(T)t_\text{Gd1} \nonumber \\
m_\text{Co1} &=  M_\text{Gd}(T)t_\text{Gd2}. \label{eq: supmag}
\end{align}
Here, $t_\text{Co1}$, $t_\text{Co2}$, $t_\text{Gd1}$, $t_\text{Gd2}$ are the thicknesses of the bottom (coming from the substrate) and top Co and Gd layers, respectively. We find the red-curve in Fig. \ref{fig:supSquid} b, which represents equation \eqref{totmag} for the stack dimensions used in the experiment, which follows the SQUID data reasonably well by choosing $M_\text{Gd0}$= 0.66 MA/m. It should be noted here that this is about a factor of 2 lower than values typically found in CoFeGd alloys \cite{Kim:2017aa}. We argue that this is a consequence of the inhomogeneous proximity induced magnetization in the Gd, which decays when moving away from the Co/Gd interface, and is therefore concentrated at the interface. In order to describe the magnetization fully, an intermixing profile and proximity-induced magnetization profile needs to be assumed and substantiated, which is beyond the scope of this work. We nonetheless choose this simple approach of homogeneous magnetization in the Gd, as the intra-layer exchange present in the Gd is expected to be strong enough to have the total angular momentum in the Gd layer respond coherently to the excitation of the DW, making the total angular momentum present in the DW the relevant parameter and not the exact way it is distributed. 

\section{All optical switching in [Co/Gd]$_2$} \label{AOSsupp}

\begin{figure}
    \centering
    \includegraphics[width=0.9\linewidth]{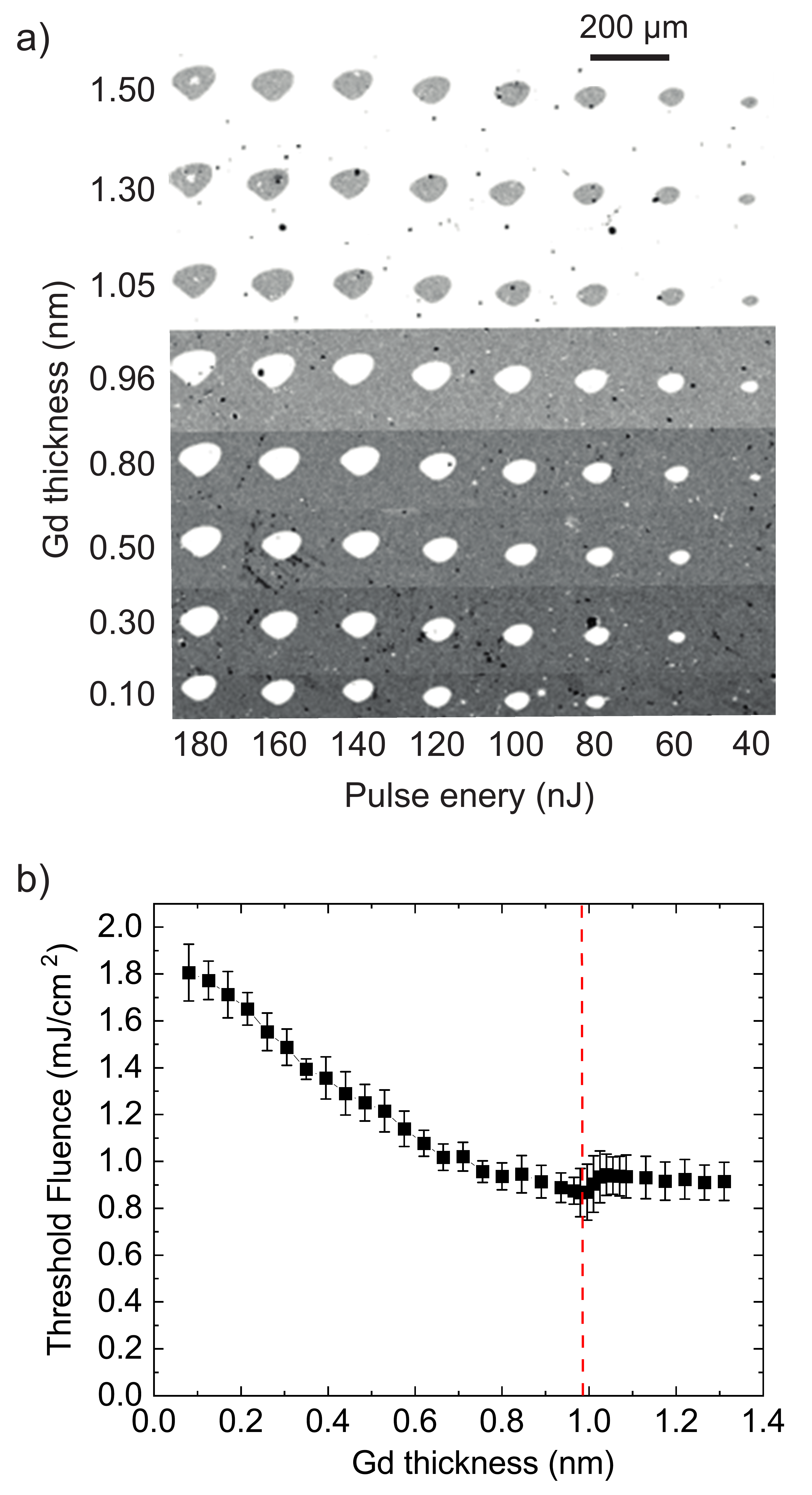}
    \caption{(a) Differential Kerr images (background taken at a Gd thickness of 0.1 nm) taken at different locations on the Gd wedge sample, which was illuminated by single laser pulses with varying pulse energy. The inverted background contrast around 1 nm is due to the transition from Co-dominated to Gd-dominated. 
    b): Calculated threshold fluence as a function of Gd layer thickness. The threshold fluence below and above compensation is calculated with the average of 4 neighboring rows.}
    \label{fig:AOS_Gd}
\end{figure}

In addition to the meta-data of all-optical switching as shown in the main text (Fig. \ref{fig:FigureStatics}.d, repeated in Fig. \ref{fig:AOS_Gd}.b), in Fig. \ref{fig:AOS_Gd}.a we present a typical excerpt of the polar Kerr microscope images used to obtain the threshold fluence as a function of Gd-thickness.

\section{Velocity enhancement expectation}\label{velocity enhancement}
To argue why a significant velocity enhancement is expected regardless of the added magnetic volume from the multiple magnetic layers, it is instructive to discuss the ratio between the expected DMI-torque-driven DW velocity in an uncompensated sample $v_\text{uc}$ to the exchange coupling torque-driven DW velocity in a compensated sample $v_\text{c}$. Here we discuss this using an analytical approach based on a 1-D model of DW motion \cite{Thiaville:2012aa, Yang:2015aa, BlaesingECT2018} (see also section \ref{Appendix: model description}), the ratio is given by:

\begin{equation}\label{ratio}
 \frac{v_{c}}{v_{uc}}=   \frac{\hbar J \theta_\text{SH} \Delta}{2e\alpha D} \frac{ \ A_\text{uc}}{ A_\text{c}}.
\end{equation}

Here, we consider the fact that $v_\text{uc}$, is limited by the strength of the DMI,
following \cite{Martinez:2014aa,Caretta:2018aa}: 
\begin{equation}
v_\text{uc} =  \frac{\pi  D}{2 \sum_{n} \mid A_n \mid}, = \frac{\pi  D}{2 A_\text{uc}},
\end{equation}
 where $D$ is the surface DMI energy density, and A$_\text{n}$ and $A_\text{uc}$ are the total angular momentum of the n$^{\text{th}}$ layer and the full stack, respectively. On the other hand, $v_\text{c}$ is for strong enough exchange coupling limited by the amount of angular momentum transferred to the DW via the spin Hall-effect, and can for a multilayered system be expressed as:
 \begin{equation}
    v_{c} =\frac{\hbar J \theta_{SHE} \Delta}{4e\alpha \sum_{n} \mid A_n \mid}=\frac{\hbar J \theta_{SHE} \Delta}{4e\alpha A_\text{c}.}, 
\end{equation}
 where $\hbar$ is the reduced Planck constant, $J$ is the current density, $\theta_\text{SHE}$ is the spin Hall angle for anti-damping-lik torque, $e$ is the electron charge, $\alpha$ is the Gilbert damping parameter, $A_\text{c}$ is the total angular momentum of the full stack, and $\Delta$ is the DW width, which shows no dependence on the DMI.
 
  To estimate the expected velocity enhancement, we consider equation \eqref{ratio} for the simple case where $A_\text{uc}/A_\text{c}=1/2$. This corresponds to the situation where one antiferromagnetically coupled layer is added to a magnetic system, which exactly compensates the angular momentum in the original system. Within this simple model, and using equation \eqref{ratio}, the condition for DW velocity enhancement then becomes:
 \begin{equation}
     \frac{v_{c}}{v_{uc}} =   \frac{\hbar J \theta_\text{SH} \Delta}{4e\alpha D} >1.
 \end{equation}

 We find that for the parameters used in the simulations (see Sup. \ref{Appendix: model description}), a net gain in DW velocity can already be obtained at a current density of 0.28 TA/m$^2$. Therefore, the exchange coupling torque in our system can be expected to result in a much higher velocity compared with the uncompensated case where the DMI-torque dominates regardless of the inevitable increase in magnetic volume for the synthetic ferrimagnets under discussion.




\section{Domain wall velocity characterization}\label{appendixsection1}

In this section, we discuss the experimental details of the characterization of the current-induced DW velocity.

We first discuss the principle of the measurement. To magnetically saturate the sample, we first reverse the magnetization in the wire with respect to the rest of the device by a combined action of an intense current pulse and an in-plane field along the current direction.  
Then, $N$ current pulses with amplitude $J$, and pulse duration $\Delta t$, displace the DW by a distance $\Delta L$, which is recorded by polar Kerr microscopy. The resulting DW velocity, $v_{DW}$, is then calculated as $v_{DW} = \frac{\Delta L}{N \Delta t}$.
We applied multiple pulses of the same amplitude to allow the DW to travel enough distance from the start to the end of the wire to minimize the errors of the length measurement and statistical errors caused by variation between pulses as a result of wire edge roughness and other potential fabrication imperfections.  
For the case of high current density or high velocity, 
this process is repeated several times for better averaging due to low $N$ being needed to translate the DW along the full wire length. 

\begin{figure*}
    \centering
    \includegraphics[width=1.0\linewidth]{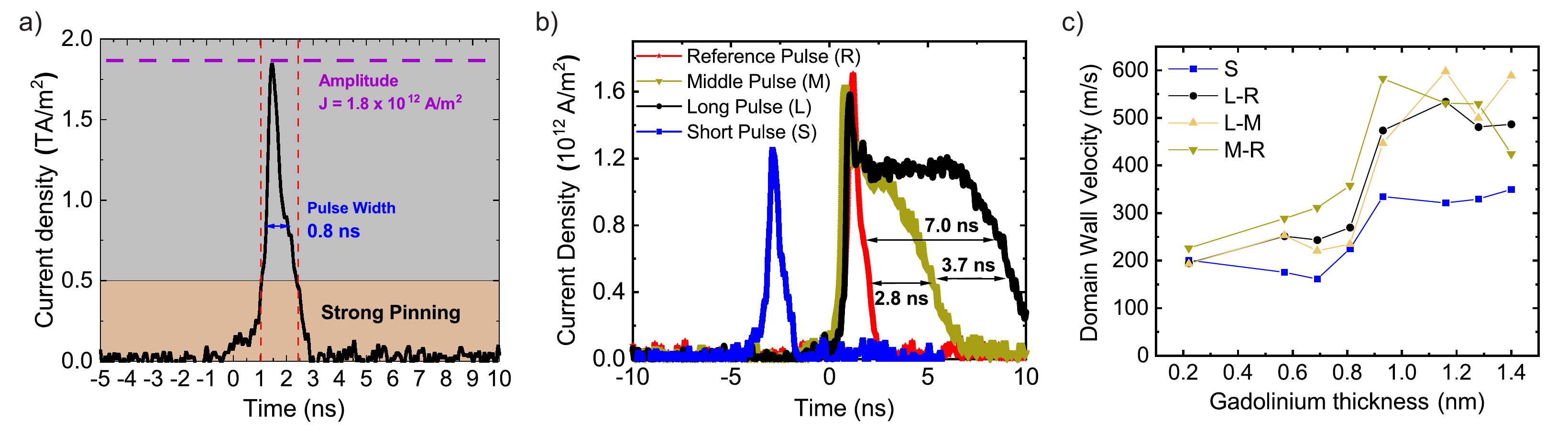}
    \caption{a): The pulse profile of a typical pulse used in the DW velocity measurements. The strong pinning region which was excluded during the pulse width estimation was marked as brown, the active region from where the effective pulse width is obtained is mark as gray. The pulse width and amplitude obtained for this characteristic pulse profile are marked in the plot. b): The pulse profile of a short pulse (S) used for the experimental results shown in the main text, a middle long pulse (M), a long pulse (L) and a reference pulse (R). The estimated time difference between those pulses are given and will be used to calculate the DW velocity. c): The DW velocity obtained from a short pulse,  and other differential pulse method. }
    \label{fig:SupDomainWall}
\end{figure*}

The method mentioned above requires the ideal case of a current pulse with a square profile. 
Next, we discuss the pulse profile used in our experiment, based on which we define our effective pulse width and amplitude. The amplitude and duration of the pulse needs to be calibrated carefully in order to clearly define the DW velocity. In our measurement set-up, we terminate the device with a 6 GHz bandwidth oscilloscope (characteristic impedance 50 $\Omega$), allowing us to characterize the electrical pulse profile directly during the measurements. 
The resulting voltage versus time profile obtained from the oscilloscope, is then converted to current density values versus time based on the DC resistance and the geometry of the devices.
Here, we have to note that in our calculations, only the thickness of Pt was considered in our calculation (assuming no current flowing in both the ferromagnetic layers and Ta). Such an assumption is made because Pt is known to be more conductive than the other metals used in the stack \cite{BlaesingECT2018}. Therefore, the energy efficiency of CIDWM in our study is a conservative estimate, since the current is likely also flowing in the other metallic layers.
To avoid dramatic heat effects induced by long current pulses, which might screen much of the physics, we used ultrashort pulses down to 1 ns in our study. The pulse amplitude can reach up to $4 \times 10^{12}$ A/m$^2$ without fully demagnetizing the sample.   
A typical measurement of the current pulse profile is shown in Fig. \ref{fig:SupDomainWall}.a. In our measurement, we define the pulse amplitude to be the peak value (1.8$\times$10$^{12}$A/m$^2$ in this case). 
The pulse exhibits a spike shape, which requires us to define an effective pulse width. We we define the effective pulse width in our measurement by taking the full width half maximum (0.8 ns) of the active area (see gray region in Fig. \ref{fig:SupDomainWall}.a) of the pulse obtained by subtracting the pulse shape by a strong, experimentally determined, pinning threshold (0.5 TA/m$^2$). Thus, the time needed for the current pulse to pull up and the decay of the pulse below the pinning threshold is not considered, since in this region the DW is not displaced or to a limited extent.
To define the pinning threshold, we applied over 500 current pulses when the DW is at various positions of the wire. The pinning threshold is then defined as the current density where no detectable movement of the DW is found. 
A further reason for using short pulses is the fact that the temperature rise during pulse time can be easily suppressed by the heat dissipation and capacity of the substrate since the heat front is still in the propagation regime. This also leads to a transient delay of temperature rise (see the next sections for its influence on the CIDWM). 

To justify the effective pulse parameters, we compare the measurement result with that obtained from a longer pulse (L) and a middle long pulse (M) shown in Fig. \ref{fig:SupDomainWall}.b, of which the shape is closer to a square pulse.  
As seen from Fig. \ref{fig:SupDomainWall}.b, the L and M pulses both have an overshoot and a slanted decay at the end of the pulse.   
We compensate for this effect by subtracting the DW displacement induced by a shorter pulse from that of a longer pulse. Since the difference between these pulse shapes resembles a square pulse shape with known pulse amplitude and width, the DW velocity can be more accurately calculated. 
However, in order to neglect the contribution from the part of the pulse that was not subtracted, it requires the thermal effect to be low (since the initial spike contributes heat as well). Thus, it is only applicable for low current density. 

Now, we introduce a reference pulse (R) (shown in red), which resemble the spike of L and M, such that its difference with L and M will results in a square pulse with a duration longer than 1 ns as shown in Fig. \ref{fig:SupDomainWall}.b. We calculated the DW velocity for the samples used in our experiment (shown in Fig. \ref{fig:DomainWallVelocity}) based on the pulse time difference between L and M (L-M), L and R (L-R), as well as M and R (M-R), which are plotted in Fig. \ref{fig:SupDomainWall}.c (current amplitude used as shown in Fig. \ref{fig:SupDomainWall}.b). We compare the obtained results with that calculated based on the short pulse (S) used in the experiment shown in the main paper (see blue curve in Fig. \ref{fig:SupDomainWall}. b), which differs in amplitude from the pulse profile in Fig. \ref{fig:SupDomainWall}. a. 
  

We observe that the DW velocities using these approaches are in the same order of magnitude, following the same general dependence on Gd thickness. The DW velocity calculated using S is found to be systematically lower.
Here we choose to be conservative instead of trying to compensate for this discrepancy, as finite thermal activation is expected to increase the DW velocity by reducing the DW friction \cite{BlaesingECT2018,Ryu:2013aa,GuanThesis}, as is the case for long pulses. So far, little study has been spent on such an issue in the flow regime of the CIDWM, although the absolute magnitude of enhancement was once shown to be 20 \% for a SAF \cite{BlaesingThesis2} for 40 K of temperature increase. So we make a compromise by taking a large but safe margin avoiding overestimation of the DW velocity as well as possible over-claim (for large velocity).   

\section{COMSOL simulations of Joule heating} \label{appendix:Joule heating}

\begin{figure*}\label{fig:SupJoule}
    \centering
    \includegraphics[width=\linewidth]{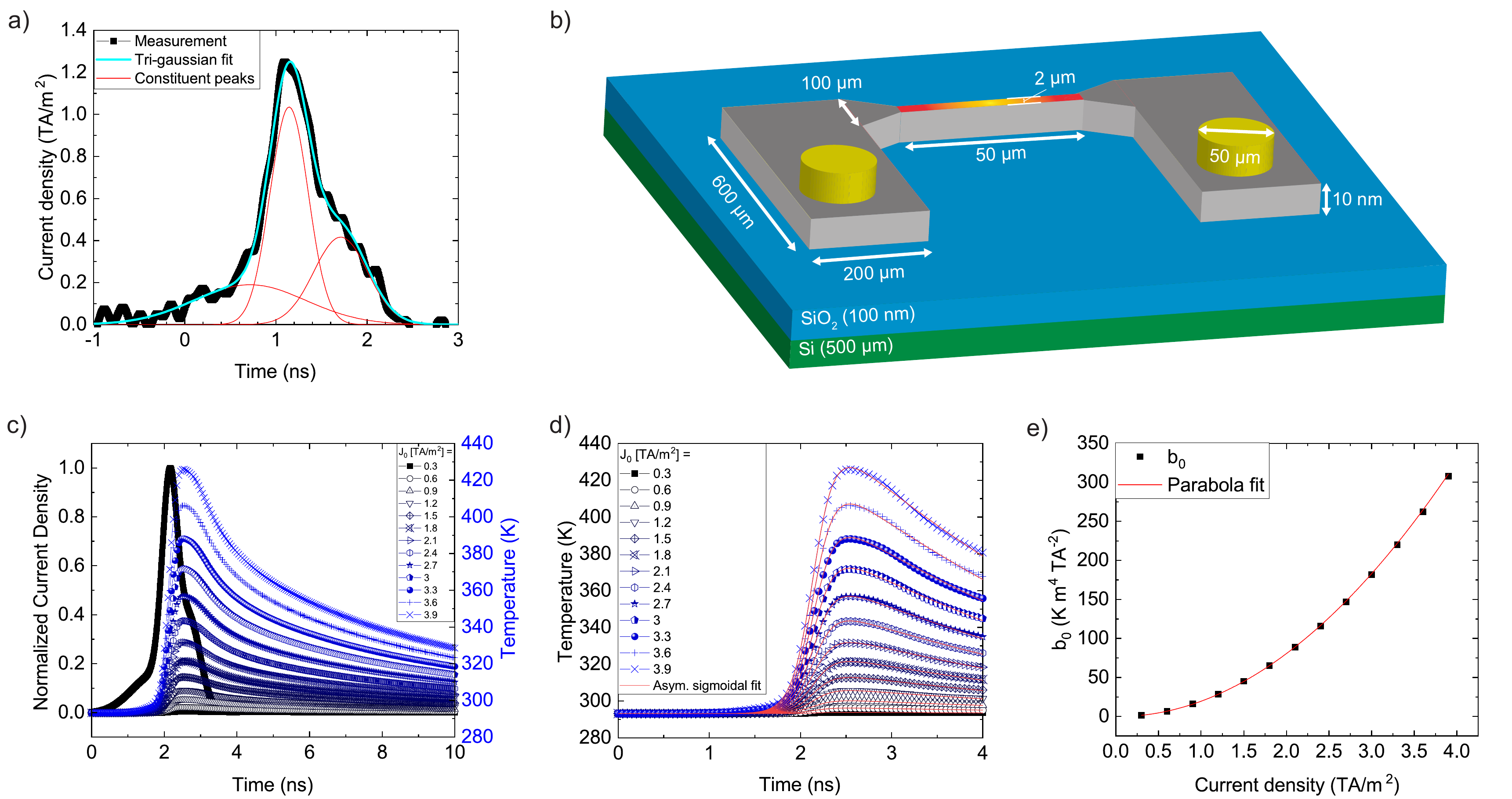}
    \caption{Joule heating analysis and current pulse definition for the model. 
    a): Fit of three Gaussians (eq. \ref{Eq: current profile}) to a typical experimental current pulse profile. b): Schematic of the device geometry used in the experiment, as well as in COMSOL.
    c): Transient temperature modeled using COMSOL for the experimental current pulse profile at various peak current densities. d): Fits of equation \ref{sigmoidal} to the temperature profile at various peak current densities $J_\text{0}$. e): Parabola fit to the prefactor $b_\text{0}$ in eq. \ref{sigmoidal} obtained from the analysis from d).}
    \label{fig:supJoule}
\end{figure*}

In this section, we present our study on the temperature rise during the application of the current pulses. This is a crucial ingredient for the analysis of the experimentally observed DW motion, as well as design considerations for future applications. Here, we based our study on COMSOL simulations, which employs multiphysics modeling including thermal transport and electrical conduction. In our model, we defined the geometry based on the real physical dimensions of the device used in this study (see Fig. \ref{fig:supJoule}.b) and the material parameters either from the COMSOL material library or from our measurements (see Table \ref{table:COMSOLpar}), which takes into account temperature-dependent effects. We model the metallic stack as a double layer consisting of a 8 nm non-conductive layer, and a 6 nm conductive layer, of which the conductivity is obtained from RT DC measurements, while the density, heat capacity and thermal conductivity of the two layers are kept the same.

\begin{table}[h]
\begin{minipage}{174pt}
\caption{Parameters used in COMSOL simulation for Joule heating induced by the current pulse.}\label{table:COMSOLpar}
\begin{tabular}{@{}lll@{}}
\toprule
\textbf{Parameter @ RT} & \textbf{Value}  &  \textbf{Unit}\\
\midrule
Thickness of the Si substrate    & 0.5   & mm   \\
Thickness of SiO$_2$     & 100   & nm  \\
Diameter of wire bond     & 50   & $\mu$m  \\
\midrule
Heat Capacity of Si Substrate    & 678   & J/(kg$\cdot$K)   \\
Density of Si    & 2320   & kg/m$^3$    \\
Thermal Conductivity of Si     & 134   & W/(m$\cdot$K)  \\
\midrule
Heat Capacity of SiO$_2$ Substrate    & 730   & J/(kg$\cdot$K)   \\
Density of SiO$_2$    & 2200   & kg/m$^3$    \\
Thermal Conductivity of SiO$_2$     & 1.4   & W/(m$\cdot$K)  \\
\midrule
Thickness of conductive metal film    & 6   & nm    \\
Thickness of non-conductive metal film    & 8   & nm    \\
Electrical Conductivity of Metal    & 0.893   & 10$^6\cdot$S/m   \\
Heat Capacity of the Metal    & 133   & J/(kg$\cdot$K)    \\
Density of the metal     & 21450   & kg/m$^3$  \\
Thermal Conductivity of the Metal    & 71.6   & W/(m$\cdot$K)   \\
Relative permittivity    & 10000   & 1    \\
\midrule
Minimum Mesh Dimension & 0.5 & nm \\
\botrule
\end{tabular}
\end{minipage}

\end{table}
\newpage
Electrical stimuli are applied through a circular region (yellow disks in Fig. \ref{fig:supJoule}.b), of which the size is close to the physical size of the electrical contacting wire bonds in our experiments. The wire is significantly heated compared with the rest of the metal structure owing to its large current density. Air convection is present only at the top surface to emulate the experimental conditions. Following these notions, we computed the temperature rise relative to RT as a result of a current pulse similar to the one used in the experiment, as indicated in Fig. \ref{fig:supJoule}.a. In order to describe this pulse profile, we fit an empirical function $J(t)$ consisting of the sum of three Gaussians: 
\begin{equation}\label{Eq: current profile}
    J(t)=\sum_{i=1}^{3}\frac{A_\text{i}}{w_\text{i}\sqrt{\pi/2}}\exp{\left(-2\frac{\left(t-t_\text{c,i}\right)^2}{w_\text{i}^2}\right)}.
\end{equation}

The best-fit parameters are summarized in table \ref{supTab2}. The same profile is used to define the voltage pulse that is applied to the system in COMSOL. The resulting temperature rise for various peak current densities is plotted in Fig. \ref{fig:supJoule}.c. One important observation here, is that the heating of the magnetic layer is delayed by approximately 0.4 ns with respect to the current pulse. Consequently, when the SOT is maximized at the peak of the current pulse, the angular momentum is still rapidly changing. Since the exchange coupling torque efficiency is linked to the angular momentum balance between the Co and Gd, this transient heating leads to a transient DW mobility.

For Joule heating, the expected temperature rise depends quadratically on the current density. In order to describe the temperature rise for any value of the peak current density $J_\text{0}$, we fit an asymmetric sigmoidial function to the temperature rise at various current densities given by:
\begin{equation}\label{sigmoidal}
    T(t)=y_\text{0}+b_\text{0}\frac{1-\exp\left(\frac{t-t_\text{c}}{l_\text{2}}\right)}{1+\exp\left(-\frac{t-t_\text{c}}{l_\text{1}}\right)}.
\end{equation}
The resulting fits for various current densities are shown in Fig. \ref{fig:supJoule}.d. During this fit all parameters except the amplitude of the peak $b_0$ were taken constant. The found amplitude, which describes the heating is then plotted as a function of current density in Fig. \ref{fig:supJoule}.e. We fit a simple quadratic function to this quantity:
\begin{equation}\label{quadratic prefactor}
b_\text{0}(J_\text{0})=k_\text{0}J_\text{0}^2,
\end{equation}
which is shown as the red line. The correspondence between the fit and the temperature rise modelled by COMSOL suggests that Joule heating plays the dominant role in our system.

In summary, based on the best fit to COMSOL simulations and the quadratic dependence of the temperature change on current density, we made an estimation of the temporal profile of the temperature rise in our sample. This asymmetric sigmoidal profile with the quadratic prefactor given in equation \ref{sigmoidal} and \ref{quadratic prefactor}, will be used to dynamically describe the change in temperature in the numerical 1-dimensional simulations of CIDWM in our [Co/Gd]$_2$ samples.
\begin{table}[h]
\begin{center}
\begin{minipage}{174pt}
\caption{Summary of the fitting parameters in equations \ref{Eq: current profile} and \ref{sigmoidal}, found to best describe the experimental current profile and the modeled temperature rise due to Joule heating, respectively.}\label{Table:CurrentPulse}%
\begin{tabular}{@{}lll@{}}
\toprule
Parameter & Value  &  Unit\\
\midrule
$t_\text{c,1}$     & -2.85   & ns  \\
$w_\text{1}$    & 0.42   & ns   \\
$A_\text{1}$    & 0.54   & TA m$^{-2}$ ns    \\
$t_\text{c,2}$     & -3.29   & ns  \\
$w_\text{2}$    & 1.23   & ns   \\
$A_\text{2}$    & 0.29   & TA m$^{-2}$ ns    \\
$t_\text{c,3}$     & -2.29   & ns  \\
$w_\text{3}$    & 0.59   & ns   \\
$A_\text{3}$    & 0.31   & TA m$^{-2}$ ns    \\
\midrule
$y_\text{0}$    & 293.15   & K   \\
$t_\text{c}$    & 2.15   & ns   \\
$l_\text{1}$    & 0.11   & ns   \\
$l_\text{2}$    & 1.94   & ns   \\
$k_\text{0}$ & 20.7 & K m$^4$ TA$^-2$\\

\botrule
\end{tabular}
\end{minipage}
\end{center}
\end{table}

\section{1-dimensional model of current induced domain wall motion}\label{Appendix: model description}
In this section the analytical 1D-model for current-induced DW motion in the [Co/Gd]$_2$ magnetic system is described. This model is an extension to earlier work by Bläsing et al. \cite{BlaesingECT2018}, who introduced the model to describe the DW dynamics of a Co/Gd bilayer. We will first summarize the main points of this model, noting that a full discussion, can be found elsewhere \cite{BlaesingThesis2}.  

The model discussed below describes the dynamics of an up-down DW in a nanowire device [length ($x$-direction) $\gg$ width ($y$-direction) $\gg$ height ($z$-direction)]. The orientation of the spins in the DW will be discussed in terms of spherical coordinates $\theta$ and $\phi$, which describe the polar and azimuthal angles of spins in the system, respectively. We define $\theta=0$ and $\theta=\pi$ to be the positive and negative out-of-plane ($z$) direction respectively, and $\phi=0$ and $\phi=\pi$ to be pointing to the positive and negative $x$-direction, respectively. Within the DW, the spins rotate gradually from $\theta=0$ to $\theta=\pi$. In equilbrium, the DW angle, which we define as the azimuthal angle of the spin in the middle of the DW will point along $\phi=\pi$ (-$x$) due to the counterclockwise DMI.

In order to then derive the equations of motion of the spins in the DW in the presence of non-conservative forces, like Gilbert damping and current-induced torques like the spin transfer torque (STT) and spin Hall effect (SHE) torque, the Rayleigh-Lagrange equation is solved:

\begin{equation}\label{eq:EulerLagrange}
    \frac{\partial L}{\partial p_\text{i}}-\frac{\text{d}}{\text{dt}}\frac{\partial L}{\partial \dot{p_\text{i}}}-\frac{\partial F}{\partial \dot{p_\text{i}}}.
\end{equation}
Here $p_\text{i}$ refers to the free variables in the system, $L$ and $F$ describe the Lagrangian and the dissipation function, respectively. Within our model of [Co/Gd]$_2$ there are five free variables: The azimuthal angles $\phi$ for the separate layers, and the DW position $q$, from which the DW velocity $\dot{q}$ is derived:
\begin{equation}
    \dot{q}=-\frac{\Delta}{\sin{\theta}}\dot{\theta}.
\end{equation}
Here it is assumed that the DW positions are tightly coupled, such that $q$ is always the same for the different layers. The quantities $L$ and $F$ are defined as:
\begin{equation}\label{eq:lagrangian}
    L= \int_{-\infty}^{\infty} \omega_\text{DW} - T \,dx ,
\end{equation}
and
\begin{equation}\label{eq:dissipation functions}
    F= \int_{-\infty}^{\infty} P_\mathrm{\alpha} + P_\mathrm{STT} + P_\mathrm{SHE} \,dx ,
\end{equation}
respectively, where $P_\mathrm{\alpha}$, $P_\mathrm{STT}$ and $P_\mathrm{SHE}$, are the dissipation functions for Gilbert Damping, STT and SHE respectively, $T$ is the "kinetic energy" of the DW which we take identical to Blaesing et al. \cite{BlaesingThesis2}, and finally $\omega_\mathrm{DW}$ is the energy density of the DW. In order to define $\omega_\mathrm{DW}$, $P_\mathrm{\alpha}$, $P_\mathrm{STT}$ and $P_\mathrm{SHE}$, it is useful to introduce some compact notation for parameters corresponding to a particular magnetic layer. We will refer to parameters corresponding to specific magnetic layers through the subscripts 1, 2, 3 and 4, where the numbering refers to the layer in the Pt/[Co/Gd]$_2$ structure starting with layer 1 from the bottom upwards (i.e. layer 1 is the Co layer interfaced with the Pt etc.). Using this notation $\omega_\mathrm{DW}$ is given by:
\begin{multline}
    \omega_\text{DW}= \frac{D_\text{1}}{\Delta}\cos{\phi_\text{1}}\sin{\theta}\\ +\sum_{i=1}^{4}K_\text{i}t_\text{i}\sin^2\theta +\sum_{i=1}^{4} \frac{ A_\text{ex,i} t_\text{i}}{\Delta^2} \sin^2\theta\\
    +\sum_{i=1}^{3} J_\text{ex} \left( \cos^2\theta - \cos\left(\phi_\text{i}-\phi_\text{i+1}\right)\sin^2\theta \right).
\end{multline}
Here, $D$ is the DMI constant, $\Delta$ is the DW width, $K_\text{i}$ is the anistropy energy density, $t_\text{i}$ is again the thickness, $A_\text{ex}$ is the exchange stiffness, and $J_\text{ex}$ is the exchange coupling strength. The DMI is assumed to only interact with the bottom Co layer ($i=1$) as it originates from the Co/Pt interface, and no DMI at the Co/Gd interface has yet been reported. The final term, describing the antiferromagnetic exchange coupling has been chosen such that negative $J_\text{ex}$ promotes antiferromagnetic coupling.

Next, the  dissipation functions are defined as:

\begin{multline}
    P_\mathrm{\alpha} + P_\mathrm{STT} + P_\mathrm{SHE}= \sum_{i=1}^{4}-\frac{\alpha_\mathrm{i},m_\mathrm{i}\Delta}{\gamma_\mathrm{i}}\left(\dot{\phi_\mathrm{i}}^2+\dot{q}^2/\Delta^2\right)\\  +\sum_{i=1}^{4} \frac{\upsilon_\mathrm{i}u_\mathrm{i} m_\mathrm{i}}{\Delta \gamma_\mathrm{i}} \sin^2\theta\dot{\phi_\mathrm{i}}-\frac{\beta_\mathrm{i}u_\mathrm{i} m_\mathrm{i}}{\Delta^2 \gamma_\mathrm{i}} \sin^2\theta\dot{q}\\
  -\frac{\hbar\theta_\mathrm{SH}J}{2 e}\sin{\theta}\left(\frac{\cos{\phi_\mathrm{1}}\dot{q}}{\Delta}+\sin{\phi_\mathrm{1}}\cos{\theta}\dot{\phi_\mathrm{1}}\right),
\end{multline}
where $\alpha$ is the Gilbert damping coefficient, $\gamma$ is the gyromagnetic ratio, $\beta$ is the parameter describing the non-adiabatic component to the STT, $\hbar$ is the reduced Planck constant, $\theta_\text{SH}$ is the spin-Hall angle, $J$ is the current density, $e$ is the electron charge, and $\upsilon_i$ is a variable that is equal to 1 and -1 in a Co layer ($i=1,3$) and Gd layer ($i=2,4$) respectively. Finally, $u_\text{i}$ is given by:

    \begin{equation}
    u_\text{i}=\frac{\hbar P_\text{i} \gamma_\text{i} d_\text{i}}{2 e m_\text{i}}J,
\end{equation}

where $P_\text{i}$ is the degree of polarization of the charge current. It should be noted that the spin-Hall current is in this work modelled to only interact with the Co layer interfaced with the Pt. Recent work has indicated that the spin coherence length in ferrimagnetic multilayers can be larger than the thickness of the individual layers in this study \cite{Yu:2019vn}. Hence, it is possible that part of the spin current injected along the z-direction can also interact with the other magnetic layers when it is (partially) transmitted through the first Co layer.

In order to then solve equation \eqref{eq:EulerLagrange} for the five degrees of freedom in our system, we evaluate the integrals in equations \eqref{eq:lagrangian} and \eqref{eq:dissipation functions} by assuming a typical Bloch profile of the polar angle $\theta$ as a function of $x$:
\begin{equation}
    \theta(x)=2 \tan^{-1}\left(\exp\left[\frac{x-q(t)}{\Delta}\right]\right).
\end{equation}
The resulting equations of motion in absence of an applied field for $q$, $\phi_\text{1}$, $\phi_\text{2}$, $\phi_\text{3}$, $\phi_\text{4}$ are then finally given by:

\begin{multline}
    \dot{q}=\frac{\Delta}{\left(\sum_{i=1}^{4} \frac{m_\mathrm{i}\alpha_\mathrm{i}}{\gamma_\mathrm{i}\Delta}\right)} \bigg(  -\sum_{i=1}^{4}\frac{m_\mathrm{i}u_\mathrm{i}\beta_\mathrm{i}}{\gamma_\mathrm{i}\Delta} \\
    - \frac{\pi \hbar \theta_\mathrm{SH} J}{4e} \cos{\phi_\mathrm{1}}
    -\sum_{i=1}^{4} \upsilon_\mathrm{i} \frac{m_\mathrm{i}\dot{\phi_\mathrm{i}}}{\gamma_\mathrm{i}} \bigg)
\end{multline}

\begin{multline}
    \dot{\phi_\mathrm{1}}=
    \frac{u_\mathrm{1}}{\alpha_\mathrm{1}\Delta}
    +\frac{D_\mathrm{1}\pi\gamma_\mathrm{1}\sin{\phi_\mathrm{1}}}{2 m_\mathrm{1}\alpha_\mathrm{1}\Delta}\\
    -\frac{J_\mathrm{ex}\gamma_\mathrm{1}\sin{\phi_\mathrm{1}-\phi_\mathrm{2}}}{m_\mathrm{1}\alpha_\mathrm{1}}+\frac{\dot{q}}{\alpha_\text{1} \Delta}
\end{multline}

\begin{multline}
    \dot{\phi_\mathrm{2}}=
    -\frac{u_\mathrm{2}}{\alpha_\mathrm{2}\Delta}
    +\frac{J_\mathrm{ex}\gamma_\mathrm{2}\sin{\phi_\mathrm{2}-\phi_\mathrm{3}}}{\alpha_\mathrm{2} m_\mathrm{2}}
    \\-\frac{J_\mathrm{ex}\gamma_\mathrm{2}\sin{\phi_\mathrm{1}-\phi_\mathrm{2}}}{\alpha_\mathrm{2} m_\mathrm{2}}+\frac{\dot{q}}{\alpha_\text{2}\Delta}
\end{multline}

\begin{multline}
    \dot{\phi_\mathrm{3}}=
    \frac{u_\mathrm{3}}{\alpha_\mathrm{3}\Delta}    -\frac{J_\mathrm{ex}\gamma_\mathrm{3}\sin{\phi_\mathrm{3}-\phi_\mathrm{2}}}{ m_\mathrm{3} \alpha_\mathrm{3}}
    \\-\frac{J_\mathrm{ex}\gamma_\mathrm{3}\sin{\phi_\mathrm{4}-\phi_\mathrm{3}}}{\alpha_\mathrm{3} m_\mathrm{3}}+\frac{\dot{q}}{\alpha_\text{3}\Delta}
\end{multline}

\begin{multline}
    \dot{\phi_\mathrm{4}}=\frac{u_\mathrm{4}}{\alpha_\mathrm{4}\Delta}
    -\frac{J_\mathrm{ex}\gamma_\mathrm{4}\sin{\phi_\mathrm{4}-\phi_\mathrm{3}}}{\alpha_\mathrm{4} m_\mathrm{4}}
    +\frac{\dot{q}}{\alpha_\text{4}\Delta}.
\end{multline}

We extended the model described earlier by Blaesing in three ways. First, we implement the thickness and temperature-dependence of the magnetization based on the SQUID measurements discussed in section \ref{section:SQUIDcharac}, in order to account for the different thicknesses of the samples investigated in the experiment. Second, the temperature of the sample, contrary to earlier work, is also taken as a dynamic parameter since we find that for the ultrashort current pulses investigated in the main text, the sample is still heating up rapidly at the peak of the current pulse, when the driving torque is largest. Finally, we trivially extended the modeled magnetic system from two to four coupled macrospins. 

\begin{table*}[h]
\begin{center}
\begin{minipage}{0.61\textwidth}
\caption{Parameters used in 1D-modeling of CIDWM in [Co/Gd]$_2$.}\label{supTab2}%
\begin{tabular*}{\textwidth}{@{}llllll@{}}
\toprule
Parameter & Unit  & Co$_\text{1}$ & Co$_\text{2}$ & Gd$_\text{1}$ & Gd$_\text{2}$ \\
\midrule
$g$     & - & 2.2$^{(a)}$  & 2.2$^{(a)}$  & 2.0$^{(b)}$  & 2.0$^{(b)}$ \\
$\gamma$     & 10$^{11}$rad/s/T  & 1.93$^{(c)}$  & 1.93$^{(c)}$   & 1.76$^{(c)}$  & 1.76$^{(c)}$  \\
$\alpha$     & - & 0.1$^{(d)}$ & 0.1$^{(d)}$  & 0.1$^{(e)}$  & 0.1$^{(e)}$ \\
$\Delta$        & nm   & 8$^{(f)}$  & 8$^{(f)}$ & 8$^{(f)}$ & 8$^{(f)}$\\
$J_\text{ex}$      & mJ/m$^2$  & -0.9$^{(g)}$  & -0.9$^{(g)}$  & -0.9$^{(g)}$  & -0.9$^{(g)}$  \\
$\beta$       & -  & -0.5$^{(h)}$ & -0.5$^{(h)}$ & -0.5$^{(h)}$ & -0.5$^{(h)}$ \\
$P$      & -  & 0.1$^{(i)}$ & 0.1$^{(i)}$ & 0.1$^{(i)}$ & 0.1$^{(i)}$ \\
$\theta_\text{SH}$    & -   & 0.08$^{(j)}$ & 0$^{(k)}$ & 0$^{(k)}$ & 0$^{(k)}$  \\
$D$   & pJ/m & 0.25$^{(l)}$ & 0$^{(m)}$ & 0$^{(m)}$ & 0$^{(m)}$ \\

\botrule
\end{tabular*}

\footnotetext[1]{From \cite{gfactorCo1,gfactorCo2}.}
\footnotetext[2]{From \cite{gfactorGd}.}
\footnotetext[3]{Calculated using the layer's respective Landé g-factor as $\gamma=\frac{g \mu_\text{B}}{\hbar}$.}
\footnotetext[4]{From \cite{Ryu:2013aa}.}
\footnotetext[5]{Assumed similar to that of Co \cite{AlphaGd}.}
\footnotetext[6]{Similar to \cite{BlaesingECT2018, Caretta:2018aa}. Assumed constant throughout the layers due to strong exchange.}
\footnotetext[7]{From \cite{ExchangeCoGd}.}
\footnotetext[8]{Estimate obtained from \cite{Okuno:2019aa}.}
\footnotetext[9]{Estimate obtained from \cite{Okuno:2019aa}.}
\footnotetext[10]{Estimate obtained from \cite{PhysRevB.96.064405,PhysRevLett.101.036601}.}
\footnotetext[11]{Spin current interaction with the 4f moments in Gd is very small \cite{ShAngleGd1,ShAngleGd2}.}
\footnotetext[12]{Chosen such that the simulations fit the $\dot{q}$ vs. $J$ experimental data.}
\footnotetext[13]{No DMI has been reported at the Co/Gd interface.}
\end{minipage}
\end{center}
\end{table*}

In order to arrive at the results presented in the main text, we numerically solve these five coupled differential equations. For this we use the temporal current density profile $J$ as given in equation \ref{Eq: current profile} and table \ref{Table:CurrentPulse}, taking into account the experimentally observed threshold current density of 0.5 TA/m$^2$ (see Sup. \ref{appendixsection1}). The magnetization and its temperature dependence are implemented as described in section \ref{section:SQUIDcharac}. All other physical parameters used in the model are summarized in table \ref{supTab2}. In order to finally characterize the velocity from the model, we divide the DW displacement by the same effective pulse width of 0.8 ns used to calculate the experimental domain wall velocity (see Sup. \ref{appendixsection1}). 

We finally note that for the realistic material parameters obtained from literature, the 1-D model generally predicts lower DW velocities than the experiments. However, an increase of the DW width from 8 to 12 nm (both reasonable for these kind of ferrimagnetic systems \cite{PhysRevB.96.064405,PhysRevLett.101.036601}) could already address this discrepancy in velocity without impacting the qualitative dependence of the DW velocity on the Gd thickness and current density.

\section{Velocity peak broadening described by dynamic temperature} \label{Motivationdyntemp}
\begin{figure*}
    \centering
    \includegraphics[width=\linewidth]{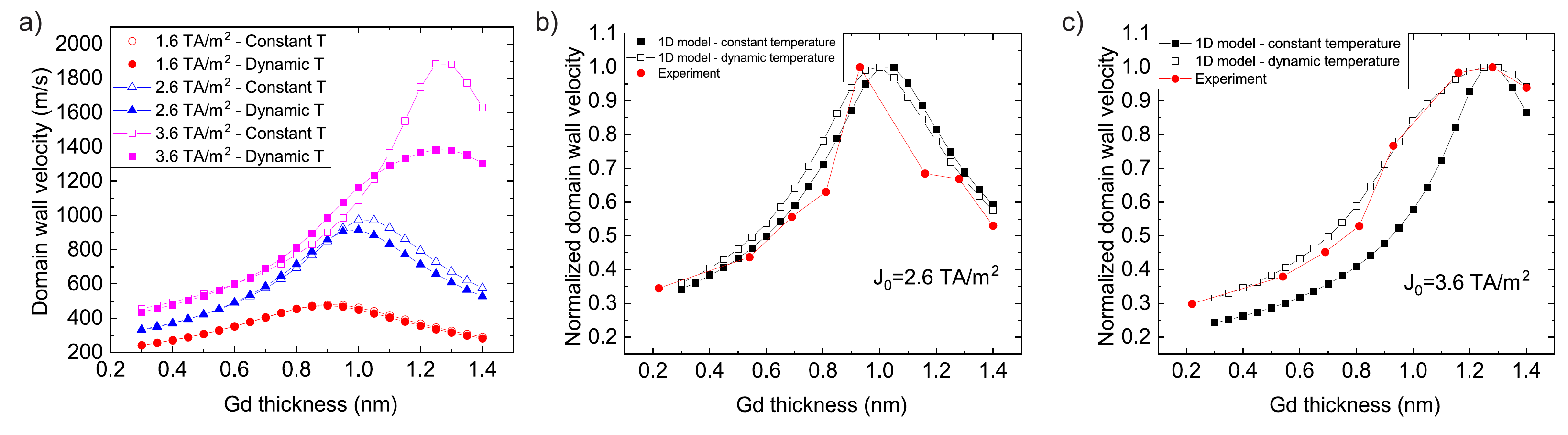}
    \caption{Comparison of using dynamic and static temperature in the 1D model of CIDWM. The constant temperature is chosen to be 85\% of the maximum temperature from the COMSOL simulations at a given peak current density, as it leads to a comparable shift of the velocity peak. a): 1D CIDWM simulations comparing the velocity for the two temperature profiles at 1.6, 2.6 and 3.6 TA/m$^2$. b): Normalized DW velocity for the 1D CIDWM simulation with the constant and dynamic temperature profiles, compared with the normalized velocities found from the experiments at peak current density $J_\text{0} = 2.6$ TA/m$^2$. c): Same as b), but for $J_\text{0} = 3.6$ TA/m$^2$.}
    \label{fig:Temperature}
\end{figure*}

To further substantiate the claim that dynamic simulation of the temperature is important to properly interpret the CIDWM experiments in our [Co/Gd]$_2$ multilayers, we make a more elaborate comparison between simulation with a dynamic temperature profile, and a constant temperature. For this we use the same simulations as presented in other parts of this work, and the same temperature profile for the dynamic temperature simulations. For the constant temperature simulations we take the temperature to be 85\% of the maximum temperature of the dynamic temperature profile, as this gives rise to a similar shift of the thickness at which the peak velocity occurs. A typical comparison of the modelled DW velocity between static and dynamic temperature can be seen in Fig. \ref{fig:Temperature} for peak current densities $J_\text{0}$ of 1.6, 2.6 and 3.6 TA/m$^2$. It can be seen that for low current densities the resulting velocity profiles are almost identical, whereas for high current densities the velocity is more sharply peaked around the compensation thickness. The main effect at high current density of the dynamic temperature profile is a simultaneous lowering of the peak velocity and a broadening of the velocity peak, both of which can be understood from the magnetic composition changing rapidly throughout the peak of the current density. As the degree of compensation changes, the resulting translation of the DW becomes more of an average over many different magnetic compositions which have varying degrees of DW mobility. 

In the experiments a similar phenomenon can be observed, and is qualitatively shown in Fig. \ref{fig:Temperature}.b and \ref{fig:Temperature}.c. Again, for the lower current density ($J_\text{0}=2.6$ TA/m$^2$, fig. \ref{fig:Temperature}.b), the heating effect within the temporal width of the current pulse is limited, and the velocity with thickness is well described by both the constant and dynamic temperature. A drastic change is observed however for large current densities ($J_\text{0}=3.6$ TA/m$^2$, fig. \ref{fig:Temperature}.c), where we find that the dynamic temperature profile describes the trend of the experiment well. Contrarily the constant temperature approximation does not work anymore beyond potentially pinpointing the shift in the velocity peak.

\end{appendices}

\bibliographystyle{ieeetr}
\bibliography{MyLib}



\end{document}